\documentclass[conference,letterpaper]{IEEEtran}
\IEEEoverridecommandlockouts % Without this \thanks{} don't appear
\usepackage[utf8]{inputenc} % The pack­age trans­lates var­i­ous stan­dard and other in­put en­cod­ings into a `LaTeX in­ter­nal lan­guage'
\usepackage[T1]{fontenc} % al­lows the user to se­lect font en­cod­ings, and for each en­cod­ing pro­vides an in­ter­face to `font-en­cod­ing-spe­cific' com­mands for each font
\usepackage[cmex10]{amsmath} % Use the [cmex10] option to ensure complicance
\usepackage{amssymb} % provides an extended symbol collection
\usepackage{mathtools} % optional corrections to amsmath, but recommended
\usepackage{amsfonts} % ex­tended set of fonts for use in math­e­mat­ics
\usepackage{accents} % A package for multiple accents in mathematics, with nice features concerning the creation of accents and placement of scripts

\usepackage{mleftright}\mleftright % defines variants \mleft and \mright of \left and \right, that make the delimiters act as \mathopen and \mathclose. These commands address spacing difficulties in subformulas.

\usepackage{multirow}
\usepackage[dvips]{graphicx} % for including graphics
\graphicspath{{./}}
\DeclareGraphicsExtensions{.eps}

\usepackage[dvipsnames]{xcolor} % provides easy driver-independent access to several kinds of colors, tints, shades, tones, and mixes of arbitrary colors; options: [xetex,dvipsnames]

\usepackage{bbm} % double-striked font for most symbols
\usepackage{xfrac} % diagonal fractions using \sfrac
\usepackage{cancel} % crossed-over equations; cancel-to-zero, etc. Options: [makeroom]
\usepackage{wrapfig} % typesetting a narrow float at the edge of the text, and making the text wrap around it.
\usepackage{tensor} % Al­lows the user to set ten­sor-style su­per- and sub­scripts with off­sets be­tween suc­ces­sive in­dices

\usepackage{theorem}
\usepackage{cite}

\usepackage{url} % The com­mand \url is a form of ver­ba­tim com­mand that al­lows line­breaks at cer­tain char­ac­ters or com­bi­na­tions of char­ac­ters, ac­cepts re­con­fig­u­ra­tion, and can usu­ally be used in the ar­gu­ment to an­other com­mand

\usepackage{psfrag} % Al­lows LaTeX construc­tions (equa­tions, pic­ture environments, etc.) to be pre­cisely superim­posed over En­cap­su­lated PostScript fig­ures

\usepackage[inline]{enumitem} % list customization; use [inline] to enable horizontal lists with *-ed environments

\usepackage{array} % extended  implementation  of  the  LaTeX array– and tabular–environments

\usepackage{aliascnt} % allows for a single counter in \autoref'ed theorems

\usepackage{hyperref} % cross-reference within document. Options: [draft,hidelinks,linktocpage]

\usepackage{cleveref} % enhances LaTeX's cross-referencing features
\usepackage[ruled,algoruled,vlined]{algorithm2e}

\SetCommentSty{mycommfont}

\usepackage{xparse} % used for conditional macro definitions

\usepackage{ifthen} % Ifthen is a sep­a­rate pack­age within the LaTeX dis­tri­bu­tion; while it will al­ways be present in a LaTeX dis­tri­bu­tion, a \usep­a­ck­age com­mand is al­ways needed to load it

\usepackage{comment}

\crefname{section}{Section}{Sections}
\crefname{subsection}{Section}{Sections}
\crefname{equation}{Eq.}{Equations}
\crefname{enumi}{part}{parts}
\crefname{table}{Table}{Tables}
\crefname{figure}{Figure}{Figures}
\crefname{algocf}{Algorithm}{Algorithms}

\newtheorem{theorem}{Theorem}
\crefname{theorem}{Theorem}{Theorems}

\newaliascnt{lemma}{theorem}
\newtheorem{lemma}[lemma]{Lemma}
\aliascntresetthe{lemma}
\crefname{lemma}{Lemma}{Lemmas}

\newaliascnt{definition}{theorem}

\aliascntresetthe{definition}
\crefname{definition}{Definition}{Definitions}

\newaliascnt{corollary}{theorem}
\newtheorem{corollary}[corollary]{Corollary}
\aliascntresetthe{corollary}
\crefname{corollary}{Corollary}{Corollarys}

\newaliascnt{claim}{theorem}

\aliascntresetthe{claim}
\crefname{claim}{Claim}{Claims}

\newaliascnt{conjecture}{theorem}

\aliascntresetthe{conjecture}
\crefname{conjecture}{Conjecture}{Conjectures}

\newaliascnt{question}{theorem}

\aliascntresetthe{question}
\crefname{question}{Question}{Questions}

\newaliascnt{oquestion}{theorem}

\aliascntresetthe{oquestion}
\crefname{oquestion}{Open Question}{Open Questions}

\theoremstyle{plain}
\theorembodyfont{\normalfont}

\newtheorem{cnstr}{Construction}%$\!$}

\newenvironment{construction}{\begin{cnstr}}{\hfill$\Box$\end{cnstr}}
\crefname{cnstr}{Construction}{Constructions}

\crefname{step}{Step}{Steps}

\crefname{regime}{Regime}{Regimes}

\newtheorem{myalgo}{Algorithm}%$\!$}

\crefname{myalgo}{Algorithm}{Algorithms}

\newcounter{enumrom}
\renewcommand{\theenumrom}{(\roman{enumrom})}

\makeatletter
\renewcommand{\@endtheorem}{\endtrivlist}
\makeatother

\makeatletter
\renewcommand{\thefigure}{{\@arabic\c@figure}}
\renewcommand{\fnum@figure}{{\bf Figure\,\thefigure}}
\makeatother

\renewcommand{\leq}{\leqslant}
\renewcommand{\geq}{\geqslant}

\newcommand{\bfc}{{\boldsymbol c}}

\newcommand{\bfu}{{\boldsymbol u}}
\newcommand{\bfv}{{\boldsymbol v}}
\newcommand{\bfw}{{\boldsymbol w}}
\newcommand{\bfx}{{\boldsymbol x}}
\newcommand{\bfy}{{\boldsymbol y}}
\newcommand{\bfz}{{\boldsymbol z}}

\newcommand{\cC}{\mathcal{C}}

\newcommand{\cF}{\mathcal{F}}

\newcommand{\cL}{\mathcal{L}}

\newcommand{\cR}{\mathcal{R}}

\newcommand{\cT}{\mathcal{T}}

\newcommand{\cX}{\mathcal{X}}

\renewcommand{\Bbb}{\mathbb}

\newcommand{\N}{{\Bbb N}}

\DeclarePairedDelimiter\abs{\lvert}{\rvert}
\DeclarePairedDelimiter\ceilenv{\lceil}{\rceil}
\DeclarePairedDelimiter\floorenv{\lfloor}{\rfloor}
\DeclarePairedDelimiter\parenv{\lparen}{\rparen}

\DeclarePairedDelimiter\bracenv{\lbrace}{\rbrace}
\DeclarePairedDelimiterX\mathset[2]{\lbrace}{\rbrace}{#1 : #2}
\DeclarePairedDelimiterX\inner[2]{\langle}{\rangle}{#1 \mathrel{},\mathrel{} #2}
\DeclarePairedDelimiterX\condparenv[2]{(}{)}{#1 \mathrel{}\delimsize\vert\mathrel{} #2}

\newcommand{\mybinom}[2]{\Biggl(\begin{array}{@{}c@{}}#1\\#2\end{array}\Biggr)}

\DeclareDocumentCommand\norm{ o m }{
    \IfNoValueTF{#1}
        {\left\Vert#2\right\Vert}
        {\left\Vert#2\right\Vert_{#1}}
}

\DeclareDocumentCommand\der{ o m o }{
    \IfNoValueTF{#1}
        {
            \IfNoValueTF{#3}
                {\frac{d}{d{#2}}}
                {\frac{d{#3}}{d{#2}}}
        }
        {\parenv*{\frac{d}{d{#2}}}^{#1}\IfNoValueTF{#3}{}{#3}}
}
\DeclareDocumentCommand\partder{ o m m }{
    \IfNoValueTF{#1}
        {\frac{\partial{#3}}{\partial{#2}}}
        {\frac{\partial^{#1}{#3}}{{\partial{#2}}^{#1}}}
}
\DeclareDocumentCommand\df{ o m o }{
    d\IfNoValueTF{#1}{}{^{#1}}{#2}\IfNoValueTF{#3}{}{_{#3}}
}

\newcommand{\deq}{\mathrel{\triangleq}}

\DeclareMathOperator{\red}{red}

\DeclareMathOperator{\rf}{\cR\cF}

\newcommand{\lmin}{L_{\min}}

\newcommand{\lover}{L_{\operatorname*{over}}}

\DeclareDocumentCommand\enc{ o }{
    \IfNoValueTF{#1}
        {\operatorname{Enc}}
        {\operatorname{Enc}_{\ref*{#1}}}
}

\DeclareDocumentCommand\dec{ o }{
    \IfNoValueTF{#1}
        {\operatorname{Dec}}
        {\operatorname{Dec}_{\ref*{#1}}}
}

\newcommand{\code}[1]{\cC_{\ref*{#1}}}

\begin{document}

\title{Reconstruction from Substrings with Partial Overlap}

%% Many authors with many affiliations:
 \author{%
   \IEEEauthorblockN{\textbf{Yonatan~Yehezkeally}\IEEEauthorrefmark{1}, 
                     \textbf{Daniella~Bar-Lev}\IEEEauthorrefmark{2}, 
                     \textbf{Sagi~Marcovich}\IEEEauthorrefmark{2}, 
                     and \textbf{Eitan~Yaakobi}\IEEEauthorrefmark{2}}
   \IEEEauthorblockA{\IEEEauthorrefmark{1}%
                     Institute for Communications Engineering, %\\
                     Technical University of Munich, 
                     80333 Munich, Germany}
   \IEEEauthorblockA{\IEEEauthorrefmark{2}%
                     Department of Computer Science, %\\
                     Technion---Israel Institute of Technology, 
                     Haifa 3200003, Israel}
   \thanks{%
   This work was supported in part by the European Research Council (ERC) through the European Union's Horizon 2020 Research and Innovation Programme under Grant 801434. 
   Y. Yehezkeally was supported by a Carl Friedrich von Siemens postdoctoral research fellowship of the Alexander von Humboldt Foundation. 
   D. Bar-Lev, S. Marcovich, and E. Yaakobi were supported in part by the United States-Israel BSF grant no. 2018048.}%
   \thanks{%
   The first three authors contributed equally to this work.}%
 }

\maketitle
\begin{abstract}
This paper introduces a new family of reconstruction codes which is motivated by applications in DNA data storage and sequencing. In such applications, DNA strands are sequenced by reading some subset of their substrings. While previous works considered two extreme cases in which \emph{all} substrings of some fixed length are read or substrings are read with no overlap, this work considers the setup in which consecutive substrings are read with some given minimum overlap. First, upper bounds are provided on the attainable rates of codes that guarantee unique reconstruction. Then, we present efficient constructions of asymptotically optimal codes that meet the  upper bound.
\end{abstract}

\section{Introduction} \label{sec:intro}

String reconstruction refers to a large class of problems where 
information about a string can only be obtained in the form of 
multiple, incomplete and/or noisy observations. 
Examples of such problems are the \emph{reconstruction problem} by 
Levenshtein~\cite{Lev01b}, the \emph{trace reconstruction 
problem}~\cite{BatKanKhaMcG04}, and the \emph{$k$-deck 
problem}~\cite{BenMeySchSmiSto91, Sco97, DudSch03, 
ChrKiaRaoVarYaaYao19}. 

Notably, when observations are composed of consecutive substrings, the 
\emph{reconstruction from substring-compositions 
problem}~\cite{MotBreTse13, BreBreTse13, AchDasMilOrlPan15, 
KiaPulMil16, MarYaa21, YehMarYaa21, AchDasMilOrlPan15, MotRamTseMa13, 
ShoCouTse15, GanMosRac16, ShoKamGovXiaCouTse16, YehPol21} and the 
\emph{torn-paper problem}~\cite{RavVahSho21, ShoVah20, NasShoVah22, 
BarMarYaaYeh22} (a problem closely related to the \emph{shuffling 
channel}~\cite{ShoHec19, HecShoRamTse17, LenSieWacYaa19, WeiMer21}) 
have received significant interest in the past decade due to 
applications in DNA- or polymer-based storage systems, resulting from 
contemporary sequencing technologies~\cite{MotBreTse13, BreBreTse13, 
GabMil19}. 
The former arises from an idealized assumption of full overlap (and 
uniform coverage) in read substrings, while the latter results from an 
assumption of no overlap; in applications, this models the question of 
whether the complete information string may be replicated and 
uniformly segmented for sequencing, or if segmentation occurs 
adversarially in the medium prior to sequencing.

Motivated by these two paradigms, we study in this paper a generalized 
(or intermediate) setting where an information string is observed 
through an arbitrary collection of its substrings, where the minimum 
length~$\lmin$ of each retrieved substring, as well as the 
length~$\lover$ of overlap between consecutive substrings, are bounded 
from below. 
A similar setting was recently studied~\cite{RavVahSho22}, where both 
substrings' lengths and overlap were assumed to be random; we study 
the problem in the aforementioned worst-case regime. 
We ask what the minimum value of $\lmin$ is for which 
there exist codes of length~$n$ strings allowing for unique 
reconstruction in this channel with asymptotically non-vanishing 
rates, and then what is the asymptotically optimal obtainable rate 
given the value of~$\lover$. Having answered both questions, we 
demonstrate that in these regimes it is possible to efficiently encode 
and decode information for unique reconstruction attaining 
asymptotically optimal rates.

The rest of this paper is organized as follows. In \cref{sec:def} we 
present notation and definitions used throughout the paper. In 
\cref{sec:partial-single} we bound from above the asymptotically 
attainable rate of codes for unique reconstruction as a function of 
$\lmin,\lover$, and then in \cref{sec:partial-single-encode} we 
develop efficient encoding and decoding algorithms for such codes, 
asymptotically meeting this bound.

\section{Definitions and Preliminaries}\label{sec:def}

Let $\Sigma$ be a finite alphabet of size $q$. Where advantageous, we 
assume $\Sigma$ is equipped with a ring structure, and in particular 
identify elements $0,1\in\Sigma$. For a positive integer $n$, let 
$[n]$ denote the set $[n] \deq \{0,1,\ldots, n-1\}$.

For two non-negative functions~$f,g$ of a common variable~$n$, 
denoting $L\deq \limsup_{n\to\infty}\frac{f(n)}{g(n)}$ (in the wide 
sense) we say that $f=o_n(g)$ if $L=0$, $f=\Omega_n(g)$ if $L>0$, 
$f=O_n(g)$ if $L<\infty$, and $f=\omega_n(g)$ if $L=\infty$. 
We say that $f=\Theta_n(g)$ if $f=\Omega_n(g)$ and $f=O_n(g)$. 
If $f$ is not positive, we say $f(n)=O_n(g(n))$ ($f(n)=o_n(g(n))$) if 
$\abs*{f(n)}=O_n(g(n))$ (respectively, $\abs*{f(n)}=o_n(g(n))$). 
If clear from context, we omit the subscript from aforementioned 
notations.

Let $\Sigma^*$ denote the set of all finite strings over $\Sigma$. 
The length of a string~$\bfx$ is denoted by $\abs*{\bfx}$. 
For strings $\bfx,\bfy\in\Sigma^*$, we denote their concatenation by 
$\bfx\circ\bfy$. We say that $\bfv$ is a \emph{substring} of $\bfx$ if 
there exist strings $\bfu,\bfw$ such that $\bfx = \bfu\circ \bfv\circ 
\bfw$. If $\bfu$ (respectively, $\bfw$) is empty, we say $\bfv$ is a 
\emph{prefix} (\emph{suffix}) of $\bfx$. If the length of $\bfv$ is 
$\ell$, we specifically say that $\bfv$ is an \emph{$\ell$-substring} 
of $\bfx$ (similarly, an $\ell$-pre/suffix). For $I\subseteq 
[\abs*{\bfx}]$, we let $\bfx_I$ denote the subsequence of $\bfx$ 
obtained by restriction to the coordinates of~$I$; specifically, for 
$i\in [\abs*{\bfx}-\ell+1]$ we denote by $\bfx_{i+[\ell]}$ the 
$\ell$-substring of $\bfx$ at \emph{location}~$i$ (we reserve the term 
\emph{index} for a different use), where $i+[\ell] = 
\mathset*{i+j}{j\in [\ell]}$ follows the standard co-set notation.

We consider in this paper the problem of string reconstruction from 
substrings with partial overlap. That is, we assume that a message 
$\bfx\in \Sigma^n$ is observed only through a multiset of its 
substrings, without order, with the following restrictions: 
\begin{enumerate*}[label=(\roman*)]
\item all observed substrings are of length at least~$\lmin$; and
\item succeeding substrings overlap with length at least~$\lover$ (in 
particular, every symbol of $\bfx$ is observed in some substring).
\end{enumerate*}

More formally, a \emph{substring-trace} of $\bfx\in\Sigma^n$ is a 
multiset $\bracenv*{\bracenv*{\bfx_{i_j+[\ell_j]}}}_{j=1}^m$, for some 
$m\in\N$, such that $i_1<i_2<\cdots<i_m$ and $\ell_j\in[n-i_j+1]$. A 
substring-trace is \emph{complete} if $i_1=0$, $i_{j+1} < i_j+\ell_j$ 
for all $j<m$, and $i_m+\ell_m-1=n$. A complete substring-trace of 
$\bfx\in\Sigma^n$ is called an \emph{$(\lmin, \lover)$-trace} if 
$\ell_j\geq \lmin\geq \lover$ for all $j\in [m]$ and $i_j + \ell_j - 
i_{j+1} \geq \lover$ for all $j\in [m-1]$. For example, for $\bfx = 
11101110101111$ 
\begin{itemize}
\item $\bracenv*{\bracenv*{1110111,111010,101111}}$ is a $(6,2)$-trace 
of $\bfx$; 

\item $\bracenv*{\bracenv*{111011,110101,101111}}$ is a complete trace 
of $\bfx$ which is not a $(6,2)$-trace; and 

\item $\bracenv*{\bracenv*{110111,110101,01111}}$ is a trace of $\bfx$ 
which is not complete. 
\end{itemize}
The \emph{$(\lmin, \lover)$-trace spectrum} of $\bfx\in \Sigma^n$, 
denoted $\cT_{\lmin}^{\lover}(\bfx)$, is the set of all $(\lmin, 
\lover)$-traces of $\bfx$. Our channel accepts $\bfx\in \Sigma^n$ and 
outputs some $(\lmin, \lover)$-trace of $\bfx$ error free. 

For all $\cC\subseteq \Sigma^n$ we denote the \emph{rate}, 
\emph{redundancy} of~$\cC$ by $R(\cC)\deq \frac{1}{n} \log\abs*{\cC}$, 
$\red(\cC)\deq n-\log\abs*{\cC}$, respectively. 
Throughout the paper, we use the base-$q$ logarithms. 
Motivated by the above channel definition, a code $\cC\subseteq 
\Sigma^n$ is called an \emph{$(\lmin, \lover)$-trace code} if for 
all $\bfx_1\neq \bfx_2\in \cC$, $\cT_{\lmin}^{\lover}(\bfx_1)\cap 
\cT_{\lmin}^{\lover}(\bfx_2) = \emptyset$. 
The main goal of this work is to find, for $\lmin,\lover$ as functions 
of~$n$, the maximum asymptotic rate of $(\lmin, \lover)$-trace 
codes. We will also be interested in efficient constructions of 
$(\lmin, \lover)$-trace codes with rate asymptotically approaching 
that value.

For convenience of analysis we denote by $\cL_{\lmin}^{\lover}(\bfx) 
\in \cT_{\lmin}^{\lover}(\bfx)$, for $\bfx\in \Sigma^n$, the 
$(\lmin,\lover)$-trace of $\bfx$ containing specifically its 
$\lmin$-prefix, and subsequent $\lmin$-substrings overlapping in 
precisely $\lover$ coordinates. For example, if $\bfx = 
11101110101111$ then 
\begin{align*}
    \cL_4^2(\bfx) = 
    \bracenv*{\bracenv*{1110,1011,1110,1010,1011,1111}}.
\end{align*}
(Here, if $\lmin-\lover$ does not divide $n-\lmin$ we allow the 
$\ell$-suffix to contain a longer overlap with its preceding 
$\ell$-substring; in the sequel we assume for ease of presentation 
that this does not occur, a fact that again shall not affect our 
asymptotic analysis.) 
Since $\cL_{\lmin}^{\lover}(\bfx)\in \cT_{\lmin}^{\lover}(\bfx)$ for 
all $\bfx\in \Sigma^n$, observe that any $(\lmin, \lover)$-trace code 
$\cC\subseteq \cX_{n,k}$ satisfies 
\begin{align}
	\abs*{\cC} &\leq 
	\abs*{\mathset*{\cL_{\lmin}^{\lover}(\bfx)}{\bfx\in \Sigma^n}}; 
\end{align}
Applying the \emph{profile-vectors} argument \cite{ChaChrEzeKia17}, we 
count the incidences of each possible $\bfu\in \Sigma^{\lmin}$ in 
$\cL_{\lmin}^{\lover}(\bfx)$ and observe that the sum of incidences 
equals $\ceilenv*{n/(\lmin-\lover)}$; thus, we have an embedding of 
$\mathset*{\cL_{\lmin}^{\lover}(\bfx)}{\bfx\in \Sigma^n}$ into 
\begin{align*}
	\mathset*{f\colon \Sigma^{\lmin}\to\N}{\sum_{\mathclap{\bfu\in 
\Sigma^{\lmin}}} f(\bfu) = \ceilenv*{\frac{n}{\lmin-\lover}}},
\end{align*}
and therefore 
\begin{align}
	\abs*{\cC} 
	&\leq \mybinom{\ceilenv*{\frac{n}{\lmin-\lover}} + q^{\lmin} - 1}{q^{\lmin} - 1} \nonumber \\
	&\leq \mybinom{\ceilenv*{\frac{n}{\lmin-\lover}} + q^{\lmin}}{q^{\lmin}}. \label{eq:code-size}
\end{align}

Before concluding this section we discuss the pertinent notion of 
\emph{repeat-free} strings~\cite{EliGabMedYaa21}, which we denote 
herein for all $\ell < n$ by 
\begin{align*}
    \rf_\ell(n)\deq 
    \mathset*{\bfx\in \Sigma^n}{\bfx_{i+[\ell]}\neq \bfx_{j+[\ell]}, 
    \forall i<j\in [n-\ell+1]}.
\end{align*}
It was observed in~\cite{Ukk1992} that for $\bfx\in \rf_\ell(n)$, 
$\cL_{\ell+1}^{\ell}(\bfx) \neq \cL_{\ell+1}^{\ell}(\bfy)$ for all 
$\bfy\in \Sigma^n\setminus\bracenv*{\bfx}$. A straightforward 
generalization of the arguments therein demonstrates the following lemma.
\begin{lemma}\label{lem:reconst-rf}
Given $\lmin>\lover$, for all $\bfx\in \rf_{\lover}(n)$, there exists 
an efficient algorithm reconstructing $\bfx$ from any 
$(\lmin,\lover)$-trace of $\bfx$.
\end{lemma}
\begin{IEEEproof}
Let $T$ be any $(\lmin,\lover)$-trace of $\bfx$. For any $\bfu\in T$, 
suppose by negation that there exist $\bfv_1, \bfv_2\in T$, $\bfv_1 \neq 
\bfv_2$, such that the $\ell_i$-suffix of $\bfv_i$ equals the 
$\ell_i$-prefix of $\bfu$, where $\ell_i\geq \lover$, for 
$i\in\bracenv*{1,2}$. Since $\bfv_1 \neq \bfv_2$, they occur in 
distinct locations in $\bfx$, and in particular their 
$\min\bracenv*{\ell_1,\ell_2}$-suffix occurs in distinct locations; 
this in contradiction to $\bfx\in \rf_{\lover}(n)$. The same argument 
proves that there do not exist $\bfv_1, \bfv_2\in T$, $\bfv_1 \neq 
\bfv_2$, such that the $\ell_i$-prefix of $\bfv_i$ equals the 
$\ell_i$-suffix of $\bfu$, where again $\ell_i\geq \lover$, for 
$i\in\bracenv*{1,2}$.

Hence, matching prefix to suffix, of lengths at least $\lover$, one 
reconstructs $\bfx$ from $T$. Equivalently, for each $\bfu\in T$, 
finding the unique $\bfv\in T$ that contains the $\lover$-prefix of 
$\bfu$ as a substring (which exists unless $\bfu$ is itself a prefix 
of $\bfx$) results with complete reconstruction. A naive 
implementation requires $O(n^2 \lover)$ run-time.
\end{IEEEproof}

Further, we note that if $\liminf \lover/\log(n) > 1$, then 
\cite{EliGabMedYaa21} showed that $\rf_{\lover}(n)$ forms a 
rate~$1-o_n(1)$ code in $\Sigma^n$ with an efficient encoder/decoder 
pair; we summarize their result in the following lemma.
\begin{lemma}\cite[Sec.~V]{EliGabMedYaa21}\label{lem:elishco}
For $k = \ceilenv*{2\log\log(n)}$ there exists an efficient encoder 
into $\rf_{\ceilenv*{\log(n)} + 5\parenv*{k+2}}(n)$, which in addition 
produces strings containing no $(2k+2)$-length run of zeros. 
Redundancy is introduced only in the initialization phase encoding 
into 
\begin{align*}
	Z(n,k)\deq \mathset*{\bfu\in \Sigma^n}{\text{$\bfu$ has no 
	length-$k$ run of zeros}}
\end{align*}
(this is the well-understood \emph{Run-length-limited (RLL)} 
constraint; see, e.g., \cite[Sec.~1.2]{MarRotSie01}). \\
Further, an efficient decoder exists for the provided encoder.
\end{lemma}

Analysis of the asymptotic rate achieved by the encoder of 
\cref{lem:elishco} is given in the following lemma.
\begin{lemma}\label{lem:rll-red}
There exist encoders into $Z(n,k)$ requiring $\ceilenv*{\frac{q}{q-2} 
n/q^k}$ redundant symbols for $q>2$ \cite[Lem.~5]{YehMarYaa21}, or 
$2\ceilenv*{2 n/2^k}$ for $q=2$ \cite[Sec.~III]{LevYaa19}.
\end{lemma}

For our purposes, however, it will be beneficial to observe that the 
arguments used in \cite[Sec.~V]{EliGabMedYaa21} apply without change 
to any other choice of $k$, resulting in the following corollary. It 
demonstrates that $\rf_{\lover}(n)$ forms a rate~$1-o_n(1)$ code in 
$\Sigma^n$ as long as $\lover - \log(n) = \omega(1)$.
\begin{corollary}\label{cor:rf}
For $\ell(n)=\ceilenv*{\log(n)}+\omega_n(1)$ and any 
\begin{align*}
	t \leq 2 \floorenv*{\parenv*{\ell(n) - \ceilenv*{\log(n)}}/5} - 3
\end{align*}
there exists an efficient encoder/decoder pair into $\rf_{\ell(n)}(n)$, 
producing strings containing no $t$-length run of zeros, and requiring 
at most $\ceilenv*{\tfrac{q^2}{q-2} q^{-\floorenv*{t/2}} n}$ redundant 
symbols for $q>2$ or $2\ceilenv*{4\cdot 2^{-\floorenv*{t/2}} n}$ 
symbols for $q=2$, i.e., rate $1-O_n(q^{-t/2})$.
\end{corollary}
\begin{IEEEproof}
For any $s\leq \floorenv*{\parenv*{\ell(n) - \ceilenv*{\log(n)}}/5} - 
2$, letting $k=s$ in \cref{lem:elishco} produces $\ell'$-repeat-free 
strings, for some $\ell'\leq \ell(n)$, hence in particular also 
$\ell(n)$-repeat-free strings, containing no
$(2s+2)$-length run of zeros. 

Letting $s\deq \floorenv*{t/2}-1$ we encode $\ell(n)$-repeat-free 
strings containing no $t$-length run of zeros. Finally, the encoders 
of \cref{lem:rll-red} then require the claimed redundancy.
\end{IEEEproof}

Given \cref{cor:rf} and the preceding discussion, we focus in the 
sequel on the complement, unsolved case of $\limsup \lover/\log(n) 
\leq 1$.

\section{Bounds}\label{sec:partial-single}

In this section we demonstrate an upper bound on the achievable 
asymptotic rate of $(\lmin, \lover)$-trace codes.

\begin{lemma}\label{lem:over-lin-rate}
	If $\lmin = a \log(n) + O_n(1)$ and $\lover = \gamma \lmin + 
	O_n(1)$, for some $a > 1$ and $0<\gamma\leq \frac{1}{a}$, then any 
	$(\lmin, \lover)$-trace code $\cC\subseteq \Sigma^n$ satisfies 
	\begin{align*}
		R(\cC) 
		\leq \frac{1-1/a}{1-\gamma} + 
		O\parenv*{\frac{\log\log(n)}{\log(n)}}.
	\end{align*}
	(Note that $\frac{1 - 1/a}{1-\gamma}\leq 1$ if and only if 
	$\lim \lover/\log(n) = \gamma a \leq 1$.)
\end{lemma}
\begin{IEEEproof}
Observe for $v\geq u\geq 0$ that 
\begin{align*}
	\log\binom{u+v}{u} &< \log\frac{(u+v)^u}{u!} 
	< \log\parenv*{\parenv*{\tfrac{e}{u}(u+v)}^u} \\
	&= u \parenv*{\log(e) + \log\parenv*{\tfrac{u+v}{u}}} \\
	&= u \parenv*{\log(e) + \log\parenv*{\tfrac{v}{u}} + \log\parenv*{\tfrac{u+v}{v}}} \\
	&\leq u \parenv*{\parenv*{1+\tfrac{u}{v}} \log(e) + \log\parenv*{\tfrac{v}{u}}} \\
	&\leq u \parenv*{2 \log(e) + \log\parenv*{\tfrac{v}{u}}}.
\end{align*}
Letting $v\deq q^{\lmin}$ and $u\deq \ceilenv*{n/(\lmin-\lover)}$, we 
observe that $\log(\frac{v}{u}) = O(\log(n))$ and hence by 
\cref{eq:code-size} we have 
\begin{align*}
	\log\abs*{\cC} 
	&\leq n \frac{\lmin - \log(n) + \log(\lmin-\lover) + 2\log(e)}{\lmin-\lover} \\
	&\phantom{=} + O(\log(n)) \\
	&= n \parenv*{\frac{\lmin - \log(n)}{\lmin-\lover} 
	+ O\parenv*{\frac{\log\log(n)}{\log(n)}}} \\
	&= n \parenv*{\frac{1-1/a}{1-\gamma} 
	+ O\parenv*{\frac{\log\log(n)}{\log(n)}}}.
\end{align*}

\vspace*{-\baselineskip}
\end{IEEEproof}

In particular, \cref{lem:over-lin-rate} implies the following lower 
bound on $\lmin$ for the existence of codes with asymptotically 
non-vanishing rates.
\begin{corollary}
If $\limsup_n \lmin/\log(n)\leq 1$, then $R(\cC)=o_n(1)$ for any 
$(\lmin, \lover)$-trace code $\cC\subseteq \Sigma^n$.
\end{corollary}

\section{Construction}\label{sec:partial-single-encode}

In this section we present an efficient encoder for 
$(\lmin, \lover)$-trace codes, achieving asymptotically optimal rate, 
for the case $\limsup \lover/\log(n) \leq 1$. Throughout the section, 
we let $\lmin\deq \ceilenv*{a \log(n)}$ and $\lover\deq 
\ceilenv*{\gamma \lmin}$, for some $a>1$ and $0 < \gamma \leq 1/a$. 
Further, we let $f$ be any function satisfying $f(n) = \omega(1)$ and 
$f(n) = o(\log(n))$, and finally $I\deq 
\ceilenv[\big]{\frac{1-\gamma a}{1-\gamma} \log(n) + 
(\log(n))^{0.5+\epsilon}}$ for some small $\epsilon>0$. In the sequel 
we tacitly assume that $q^I \lmin$ divides~$n$.

\begin{construction}\label{cnst:overlap}
The construction is based on the following two ingredients:

\begin{itemize}
\item \emph{Index generation}. 
Let $\parenv*{\bfc_i}_{i\in[q^I]}$, $\bfc_i\in\Sigma^I$ be indices in 
ascending lexicographic order. We encode each $\bfc_i$ independently 
as follows (see~\cref{fig:index_gen}). 
Denoting $F\deq \ceilenv*{I/f(n)}-1$, we partition $\bfc_i$ into 
$F+1$ non-overlapping segments; more formally, define 
$\bracenv*{\bfc_i^{(k)}}_{k\in [F]}\subseteq \Sigma^{f(n)}$, and 
$\bfc_i^{(F)}\in \Sigma^{I \bmod f(n)}$, by 
\begin{align*}
	\bfc_i^{(0)}\circ \bfc_i^{(1)}\circ \cdots\circ \bfc_i^{(F)} 
	= \bfc_i.
\end{align*}
Lastly, denote $\bfc'_i{}^{(k)}\deq 1\circ \bfc_i^{(k)}\circ 1\in 
\Sigma^{f(n)+2}$ (similarly, $\bfc'_i{}^{(F)}$). We refer to $\bfc_i$ 
(or simply~$i$) as an \emph{index} in the construction, and to 
$\bracenv*{\bfc'_i{}^{(k)}}_{k\in [F+1]}$ as segments of an 
\emph{encoded index}.

\begin{figure}[t]{}%
\centering
\psfrag{parity symbol}{parity symbol}
\includegraphics[width=0.95\columnwidth]{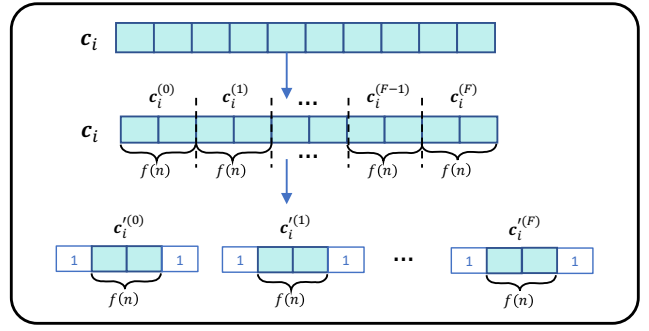}%
\caption{Index generation. Each index $\bfc_i$ is first partitioned 
into $F+1$ non-overlapping segments of length $f(n)$. Then, each of 
the segments is concatenated with a single $1$ in each edge.   
\label{fig:index_gen}}
\end{figure}

\item \emph{Repeat-free (RF) encoding}. 
For $\ell\in\N$, we use an RF encoder which receives strings of 
length~$m$ and returns strings of length $N_{n,\ell}(m)$ which contain 
no repeated substrings of length~$\ell$, and no zero-runs of 
length~$f(n)$ (see \cref{cor:rf}). We denote such an encoder by 
$E^{\rf}_{m,\ell}$.
\end{itemize}

The constructed code, denoted by $\code{cnst:overlap}(n)$, is 
carried as follows. Denote 
\begin{align*}
	r &\deq f(n)+3 + (F+1)(f(n)+2), \\
	\ell &\deq \ceilenv[\bigg]{\frac{\lover-2f(n)-5}{1 + 
	(f(n)+2)\big/\floorenv[\big]{\frac{\lmin-r}{F+1}}}}.
\end{align*}

Let $m$ be an integer such that 
\begin{align*}
	N_{n,\ell}(m) = q^{-I} n \parenv*{1 - r/\lmin}.
\end{align*}
(See the proof of \cref{thm:single-red} for an explanation of why 
such~$m$ exists for the choices of $n,\lmin,f(n),I,r,\ell$.)
An information string $\bfx\in \Sigma^{q^Im}$ is first partitioned 
into $q^I$ non-overlapping segments $\bfx = \bfx_0\circ \cdots \circ 
\bfx_{q^I-1}$, where $\bfx_i\in \Sigma^m$ for each $i\in [q^I]$. Each 
$\bfx_i$ is then independently encoded into a string $\bfz_i$ of length~$q^{-I}n$ as explained below; the motivation for this step, 
which we later describe more formally, is to satisfy two properties: 
\begin{enumerate*}[label=(\roman*)]
\item the index~$i$ can be decoded from any $\lmin$-substring of 
$\bfz_i$; and 
\item the string $\bfz_i$ can be uniquely reconstructed from an 
$(\lmin, \lover)$-trace of $\bfz_i$.
\end{enumerate*}
Lastly, we let
\begin{align*}
	\enc[cnst:overlap](\bfx)\deq 
	\bfz = \bfz_0\circ \cdots\circ \bfz_{q^I-1}.
\end{align*}

The encoding $\bfx_i\mapsto \bfz_i$, for all $i\in [q^I]$, is 
performed as follows (see~\cref{fig:x_i encoding}). 

\begin{enumerate}
\item
Let $\bfy_i\deq E^{\rf}_{m,\ell}(\bfx_i)$, where 
$\bfy_i\in\Sigma^{N_{n,\ell}(m)}$. 

\item
Partition $\bfy_i$ into $n/(q^I \lmin)$ non-overlapping segments of length~$\lmin-r$ (recall $\parenv*{\lmin - r} n/(q^I \lmin) = N_{n,\ell}(m)$) by denoting 
\begin{align*}
	\bfy_i 	= \bfy_{i,0}\circ \bfy_{i,1}\circ \cdots\circ 
	\bfy_{i,n/(q^I \lmin)-1}.
\end{align*}

\item\label{step:index-injection}
For all $j\in [n/(q^I \lmin)]$:
\begin{enumerate}
	\item
	Partition each $\bfy_{i,j}$ into $F+1$ non-overlapping segments of 
	equal lengths (up to $\pm 1$, if necessary)
	\begin{align*}
		\bfy_{i,j} = \bfy_{i,j}^{(0)}\circ \bfy_{i,j}^{(1)}\circ 
		\cdots\circ \bfy_{i,j}^{(F)}.
	\end{align*}
	
	\item
	Combine $\mathset[\big]{\bfy_{i,j}^{(k)}}{k\in [F+1]}$ with 
	segments of the encoded index~$i$, as follows. Define for all 
	$k\in [F+1]$ 
	\begin{align*}
		\bfz_{i,j}^{(k)} \deq \bfy_{i,j}^{(k)}\circ \bfc'_i{}^{(k)}, 
	\end{align*}
	then 
	\begin{align*}
		\bfz_{i,j}\deq \begin{cases}
		1 0^{f(n)} 1 1\circ \bfz_{i,j}^{(0)}\circ \cdots
		\circ \bfz_{i,j}^{(F)}, & j=0; \\
		1 0^{f(n)} 0 1\circ \bfz_{i,j}^{(0)}\circ \cdots
		\circ \bfz_{i,j}^{(F)}, & j>0
		\end{cases}
	\end{align*}
	(where we refer to the substrings $1 0^{f(n)} 0 1, 1 0^{f(n)} 1 1$ 
	as \emph{synchronization markers}).
\end{enumerate}

\item
Concatenate 
\begin{align*}
	\bfz_i\deq \bfz_{i,0}\circ \cdots\circ \bfz_{i,n/(q^I \lmin)-1}.
\end{align*}
\end{enumerate}

\begin{figure}[t]{}%
\centering
\includegraphics[width=0.95\columnwidth]{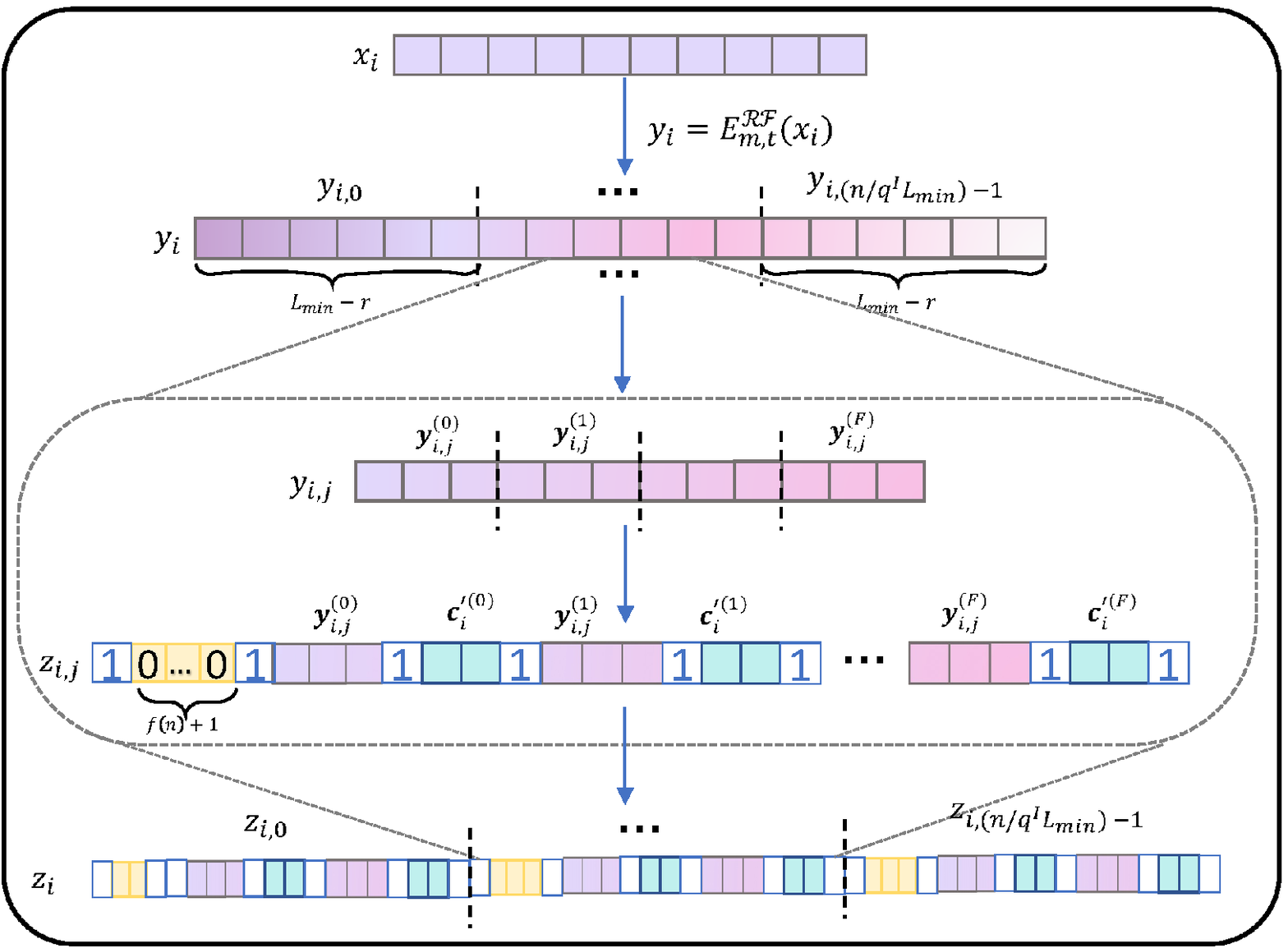}%
\caption[Encoding]{Encoding $\bfx_i$ into $\bfz_i$, as detailed in 
\cref{cnst:overlap}.
\label{fig:x_i encoding}}
\end{figure}
\end{construction}

Before we prove the correctness of \cref{cnst:overlap}, we first 
analyze $R(\code{cnst:overlap})$.
\begin{lemma}\label{lem:overlap-prop}
\begin{enumerate}
\item \label{it:cnst-red}
$r = I + \frac{2 I}{f(n)} + 2 f(n) + O(1)$.

\item \label{it:cnst-over}
Denoting $\lambda\deq 1-\frac{I}{\lmin}$, we have $\ell = \lambda 
\lover - O(f(n))$.
\end{enumerate}
\end{lemma}
\begin{comment}
\begin{IEEEproof}
\begin{enumerate}
\item 
We note 
\begin{align*}
	r &= f(n)+3 + (F+1)(f(n)+2) \\
	&= 2 f(n) + I \tfrac{f(n)+2}{f(n)} + O(1)\\
	&= I + \tfrac{2I}{f(n)} + 2 f(n) + O(1).
\end{align*}

\item
First, observe 
\begin{align*}
	(F+1)(f(n)+2) = I 
	\parenv*{1 + O\parenv*{\frac{f(n)}{\log(n)} + \frac{1}{f(n)}}}.
\end{align*}
Hence, 
\begin{align*}
	\frac{f(n)+2}{\floorenv[\big]{\frac{\lmin-r}{F+1}}} 
	&= \frac{(F+1) (f(n)+2)}{\lmin - r - O(F)} \\
	&= \frac{I \parenv*{1 + O\parenv*{\frac{f(n)}{\log(n)} + \frac{1}{f(n)}}}}{(\lmin - I) \parenv*{1 - O\parenv*{\frac{1}{f(n)} + \frac{f(n)}{\log(n)}}}} \\
	&= \frac{1-\lambda}{\lambda} + O\parenv*{\frac{f(n)}{\log(n)} + \frac{1}{f(n)}}.
\end{align*}
Finally, 
\begin{align*}
	\ell &= \lambda \frac{\lover - 2 f(n) - 5}{1 + 
	O\parenv*{\frac{f(n)}{\log(n)} + \frac{1}{f(n)}}} + O(1) \\
	&= \lambda \lover - O(f(n)).
\end{align*}
\end{enumerate}
\end{IEEEproof}
\end{comment}

Based on these properties, we have the following theorem. 
\begin{theorem}\label{thm:single-red}
Letting $f(n)\deq \ceilenv*{\sqrt{\log(n)}}$ we have 
\begin{align*}
	\red\parenv*{\cC_{\ref*{cnst:overlap}}(n)} 
	\leq n \Bigg(\frac{1/a-\gamma}{1-\gamma} &+ 
	\frac{1/a}{(\log(n))^{0.5-\epsilon}} \\
	&+ O\parenv*{\frac{1}{\sqrt{\log(n)}}}\Bigg).
\end{align*}
\end{theorem}
\begin{IEEEproof}
Observe that 
\begin{align*}
	\red\parenv*{\cC_{\ref*{cnst:overlap}}(n)} 
	&= q^I \parenv*{(N_{n,\ell}(m) - m) + \frac{n}{q^I \lmin} r}.
\end{align*}

Recalling $q^I N_{n,\ell}(m) = \parenv*{1 - \frac{r}{\lmin}} n$, we 
observe 
\begin{align*}
	\log(N_{n,\ell}(m)) &= \log(n)-I + \log\parenv*{1-\frac{r}{\lmin}} \\
	&= \frac{(a-1) \gamma}{1-\gamma} \log(n) - 
	(\log(n))^{0.5+\epsilon} + O(1)
\end{align*}
whereas by \cref{it:cnst-over} of \cref{lem:overlap-prop} 
\begin{align*}
	\ell &= \parenv*{1 - \frac{1/a-\gamma}{1-\gamma} - 
	\frac{(\log(n))^\epsilon}{a \sqrt{\log(n)}}} \gamma a \log(n) - 
	O\parenv*{\sqrt{\log(n)}} \\
	&= \frac{(a-1) \gamma}{1-\gamma} \log(n) - 
	\gamma (\log(n))^{0.5+\epsilon} - O\parenv*{\sqrt{\log(n)}}.
\end{align*}
Hence 
\begin{align*}
	\ell-\log(N_{n,\ell}(m)) = (1-\gamma) (\log(n))^{0.5+\epsilon} - 
	O\parenv*{\sqrt{\log(n)}}
\end{align*}
and for sufficiently large~$n$, 
\begin{align*}
	f(n) \leq 2 \floorenv*{\parenv*{\ell - 
	\ceilenv*{\log(N_{n,\ell}(m))}}/5} - 3,
\end{align*}
satisfying the condition of \cref{cor:rf}. We can therefore 
efficiently encode $\bfy_i = E^{\rf}_{m,\ell}(\bfx_i)$ (and vice 
versa, decode $\bfx_i$) while attaining 
\begin{align*}
	N_{n,\ell}(m) - m &\leq O\parenv*{q^{-f(n)/2} N_{n,\ell}(m)}.
\end{align*}
It follows that 
\begin{align*}
	\red\parenv*{\cC_{\ref*{cnst:overlap}}(n)} 
	\leq q^I + 
	n \parenv*{\frac{r}{\lmin} + O\parenv*{q^{-f(n)/2}}}.
\end{align*}
By \cref{it:cnst-red} of \cref{lem:overlap-prop}, 
\begin{align*}
	\frac{r}{\lmin} = \frac{1/a-\gamma}{1-\gamma} + 
	\frac{1/a}{(\log(n))^{0.5-\epsilon}} + 
	O\parenv*{\frac{1}{\sqrt{\log(n)}}},
\end{align*}
satisfying the claim.
\end{IEEEproof}
We note that the choice $f(n) = \ceilenv*{\sqrt{\log(n)}}$ in 
\cref{thm:single-red} is optimal, since $\epsilon$ in 
\cref{cnst:overlap} must satisfy $\epsilon\geq 
\max\bracenv*{\frac{\log(f(n))}{\log\log(n)}, 
1-\frac{\log(f(n))}{\log\log(n)}}-0.5$.

Finally, we prove the correctness of \cref{cnst:overlap}. We begin with two technical lemmas.
\begin{lemma}\label{lem:suff-pre-pair}
Every $\lmin$-substring $\bfu$ of $\bfz$ contains as subsequences at 
least an $(I-\mu)$-suffix of $\bfc_i$, and an $\mu$-prefix of either 
$\bfc_i$ or $\bfc_{i+1}$, for some $i\in [q^I]$ and $\mu\in [I]$, in 
identifiable locations.
\end{lemma}
\begin{IEEEproof}
For all $i\in [q^I]$ and $j\in [n/(q^I \lmin)]$, observe 
$\abs*{\bfy_{i,j}} = \lmin-r$ by construction. Since 
$\abs*{\bfc'_i{}^{(k)}}\leq f(n)+1$
for all $k\in [F+1]$, we have 
by \cref{step:index-injection} of \cref{cnst:overlap} that 
$\abs*{\bfz_{i,j}}\leq \abs*{\bfy_{i,j}} + (f(n)+3) + 
(F+1)(f(n)+2) = \abs*{\bfy_{i,j}} + r = \lmin$.

Next, observe that instances of synchronization markers only appear in 
$\bfz$ at the beginning of $\bracenv*{\bfz_{i,j}}_{i,j}$. From the 
last paragraph, either $\bfu$ contains a complete synchronization 
marker as substring, or it contains a suffix-prefix pair whose 
concatenation is an instance of a synchronization marker; in both 
cases, the exact locations in which symbols of the indices 
$\bracenv*{\bfc'_i{}^{(k)}}$ appear can be determined. 
Extracting $\bracenv*{\bfc_i{}^{(k)}}$, these contain a suffix of 
$\bfc_i$ and a prefix of either $\bfc_i, \bfc_{i+1}$ (depending on 
whether $\bfu$ is a substring of $\bfz_i$ for some~$i$) whose combined 
lengths is $I$, again since for all~$i,j$, $\abs*{\bfz_{i,j}}\leq 
\lmin$ and $\bfz_{i,j}$ contains all symbols of $\bfc_i$. 
Taking $\mu\in [I]$ to be the length of the prefix ($\mu = 0$ 
indicates the possibility that all symbols of the same index appear 
in~$\bfu$) concludes the proof.
\end{IEEEproof}

\begin{lemma}\label{lem:consec}
Every $\lover$-substring $\bfv$ of $\bfz$ contains at least~$\ell$ 
consecutive symbols of $\bfy\deq \bfy_0\circ \cdots\circ 
\bfy_{q^I-1}$.
\end{lemma}
\begin{IEEEproof}
At worst, $\bfv$ either begins or ends with a complete instance of a 
synchronization marker; hence the remaining $\lover-f(n)-3$ symbols 
are sampled from the $\bfz_{i,j}^{(k)}$ segments, and again, at worst 
end with a complete segment of an encoded index. Since 
$\abs*{\bfc'_i{}^{(k)}}\leq f(n)+2$ and $\abs[\big]{\bfy_{i,j}^{(k)}}
\geq \floorenv*{\frac{\lmin-r}{F+1}}$ for all $i,j,k$, $\bfv$ contains 
at least 
\begin{align*}
	\ceilenv[\bigg]{\frac{\lover-2f(n)-5}{1 + 
	(f(n)+2)\big/\floorenv[\big]{\frac{\lmin-r}{F+1}}}} = \ell
\end{align*}
consecutive symbols of $\bfy$.
\end{IEEEproof}

Combining both lemmas, we have the following theorem.
\begin{theorem}
For all admissible values of $n$, the code $\code{cnst:overlap}(n)$ is 
an $(\lmin, \lover)$-trace code.
\end{theorem}
\begin{IEEEproof}
Take $\bfz\in \code{cnst:overlap}(n)$ and let $T\in 
\cT_{\lmin}^{\lover}(\bfz)$ be any $(\lmin, \lover)$-trace of $\bfz$.

For $\bfu\in T$, we extract the $(I-\mu)$-suffix of $\bfc_i$, and an 
$\mu$-prefix of either $\bfc_i$ or $\bfc_{i+1}$, for some $i$, 
guaranteed by \cref{lem:suff-pre-pair}. 
Observe that if this prefix belongs to $\bfc_{i+1}$, then $\bfu$ 
also contains a complete synchronization marker $1 0^f(n) 1 1$ (the 
instance appearing as prefix of $\bfz_{i+1}$), hence these two cases 
may be distinguished. Further, note that the $\mu$-prefix of 
$\bfc_{i+1}$ equals the $\mu$-prefix of $\bfc_i$, unless the 
$(I-\mu)$-suffix of $\bfc_i$ is the greatest element of 
$\Sigma^{I-\mu}$ (in lexicographic order). In both cases, one can 
correctly deduce that the location of $\bfu$ in $\bfz$ begins in the 
segment $\bfz_i$. It is therefore possible to partition $T$ by 
index~$i$ (corresponding to the starting location of each substring).

For each substring $\bfu$ of index~$i$, intersecting both $\bfy_i, 
\bfy_{i+1}$, $\bfu$ must contain a complete synchronization marker 
$1 0^f(n) 1 1$ (the instance appearing as prefix of $\bfz_{i+1}$); 
hence its location in $\bfu$ implies the exact location of $\bfu$ in 
$\bfz$. 
For all other substrings of index~$i$, it holds by \cref{lem:consec}, 
and since each $\bfy_i$ is $\ell$-repeat-free, that there exist a 
unique way to concatenate these substrings (excluding overlap) as 
shown in \cref{lem:reconst-rf}.

Finally, once $\bfz$ is reconstructed we may extract 
$\bracenv*{\bfy_i}_{i\in [q^I]}$, then decode 
$\bracenv*{\bfx_i}_{i\in [q^I]}$ with the decoder of 
$E^{\rf}_{m,\ell}$.
\end{IEEEproof}

%%%%%%%%%%%%%%%%%%%%%%%%%%%%%%%%%%%%%%%%%%%%%%%%%%%%%%%%%%%%%%%%%%%%%%
%%%%%%%%%%%%%%%%%%%%%%%%%%%%%%%%%%%%%%%%%%%%%%%%%%%%%%%%%%%%%%%%%%%%%%
%\section*{Acknowledgments}
%
%The authors gratefully acknowledge the two anonymous reviewers and 
%associate editor, whose insight and suggestions helped shape 
%this paper.

%%%%%%%%%%%%%%%%%%%%%%%%%%%%%%%%%%%%%%%%%%%%%%%%%%%%%%%%%%%%%%%%%%%%%%
%%%%%%%%%%%%%%%%%%%%%%%%%%%%%%%%%%%%%%%%%%%%%%%%%%%%%%%%%%%%%%%%%%%%%%
%%%%%%%%%%%%%%%%%%%%%%%%%%%%%%%%%%%%%%%%%%%%%%%%%%%%%%%%%%%%%%%%%%%%%%
%%%%%%%%%%%%%%%%%%%%%%%%%%%%%%%%%%%%%%%%%%%%%%%%%%%%%%%%%%%%%%%%%%%%%%
%%%%%%%%%%%%%%%%%%%%%%%%%%%%%%%%%%%%%%%%%%%%%%%%%%%%%%%%%%%%%%%%%%%%%%
%%%%%%%%%%%%%%%%%%%%%%%%%%%%%%%%%%%%%%%%%%%%%%%%%%%%%%%%%%%%%%%%%%%%%%
%%%%%%%%%%%%%%%%%%%%%%%%%%%%%%%%%%%%%%%%%%%%%%%%%%%%%%%%%%%%%%%%%%%%%%
%%%%%%%%%%%%%%%%%%%%%%%%%%%%%%%%%%%%%%%%%%%%%%%%%%%%%%%%%%%%%%%%%%%%%%
%\bibliographystyle{IEEEtranS}
%\bibliography{allbib}
%%%%%%%%%%%%%%%%%%%%%%%%%%%%%%%%%%%%%%%%%%%%%%%%%%%%%%%%%%%%%%%%%%%%%%
%%%%%%%%%%%%%%%%%%%%%%%%%%%%%%%%%%%%%%%%%%%%%%%%%%%%%%%%%%%%%%%%%%%%%%
% Generated by IEEEtranS.bst, version: 1.14 (2015/08/26)

%%%%%%%%%%%%%%%%%%%%%%%%%%%%%%%%%%%%%%%%%%%%%%%%%%%%%%%%%%%%%%%%%%%%%%
%%%%%%%%%%%%%%%%%%%%%%%%%%%%%%%%%%%%%%%%%%%%%%%%%%%%%%%%%%%%%%%%%%%%%%

%\appendix

%%%%%%%%%%%%%%%%%%%%%%%%%%%%%%%%%%%%%%%%%%%%%%%%%%%%%%%%%%%%%%%%%%%%%%
%%%%%%%%%%%%%%%%%%%%%%%%%%%%%%%%%%%%%%%%%%%%%%%%%%%%%%%%%%%%%%%%%%%%%%

\end{document}